# Non-Volatile Magnonic Logic Circuits Engineering

Alexander Khitun and Kang L. Wang

Device Research Laboratory, Electrical Engineering Department, University of California

Los Angeles, Los Angeles, California, 90095

## Abstract

We propose a concept of magnetic logic circuits engineering, which takes an advantage of magnetization as a computational state variable and exploits spin waves for information transmission.  The circuits consist of magneto-electric cells connected via spin wave buses.  We present the result of numerical modeling showing the magneto-electric cell switching as a function of the amplitude as well as the phase of the spin wave. The phase-dependent switching makes it possible to engineer logic gates by exploiting spin wave buses as passive logic elements providing a certain phase-shift to the propagating spin waves. We present a library of logic gates consisting of magneto-electric cells and spin wave buses providing 0 or π phase shifts. The utilization of phases in addition to amplitudes is a powerful tool which let us construct logic circuits with a fewer number of elements than required for CMOS technology. As an example, we present the design of the magnonic Full Adder Circuit comprising only 5 magneto-electric cells. The proposed concept may provide a route to more functional wave-based logic circuitry with capabilities far beyond the limits of the traditional transistor-based approach.

**I. Introduction**

Modern computing devices are based on integrated circuits consisting of a large number of transistors manufactured on the surface of a silicon substrate. The transistors are connected via metallic wires in a certain way to provide logic correlation between the input and output voltages. In the past seven decades, a strait-forward approach to computation power enhancement was associated with the increase of the number of transistors per chip area and speeding up the switching of the individual transistor. According to the Moor's law[1], the number of transistors per integrated circuit doubled every eighteen months since 1958, while the switching frequency increased from kHz to the GHz range. Both these trends lead to the increase of the dissipation power, which has emerged as one of the main challenges on the way for further computational throughout increase. The power dissipation problem becomes critical with scaling down the gate length of the transistor to the nanometer range due to the quantum mechanical effects drastically increasing leakage currents. The increasing number of interconnects is another problem limiting the performance of modern computing devices. Today, it takes seven layers of metallic wires to interconnect one layer of transistors. Joule heat losses and RC delays in the interconnects mainly define the overall logic circuit performance. These problems stimulate a great deal of interest to novel materials, devices and computational paradigms able to overcome the constraints inherent to the complementary metal-oxide-semiconductor (CMOS)-based circuitry and provide a route to more functional and less power-consuming logic devices.

So far, the most of the research has been focused on the development of a novel electronic switch, which can be faster and less energy consuming than the silicon Field Effect Transistor. There is still some room for the transistor-based technology improvement by utilizing novel materials (e.g. graphene-based electronics [2]). The latter may provide some extension to the conventional approach, although without a feasible long time pathway for Moor's law continuation. A radical solution would be the development of computational paradigms able to provide a fundamental advantage over the transistor-based approach. There are some constrains inherent to the transistor-based logic circuitry: (i) the computational state variable is a scalar quantity (voltage); (ii) the metallic interconnects do not have any functional work in terms of modulating the transmitting electric signals; (iii) the transistor-based approach is volatile, requiring a permanent power supply even no computation is performed. Addressing these issues is the key leading to more efficient logic circuitry. It would be of great importance to utilize vector state variable (e.g. magnetization) allowing for amplitude as well as for phase-dependent switching. The latter will make it possible to exploit the interconnecting wires as passive logic elements modulating the phases of the transmitted signals. The incorporation of non-volatile memory elements within the body of the data processing circuit would eliminate the need in of static power consumption. All together, it may open a new horizon for logic circuitry with capabilities far beyond the limits of the CMOS technology.

As a possible solution, we consider magnonic logic circuits consisting of magneto-electric elements connected via spin wave buses. In our preceding works [3-5], we have developed the general concept of spin wave logic devices. The basic idea is to use magnetic films as spin conduit of wave propagation or referred to as – spin wave bus, where the information can be coded into a phase of the propagating spin wave. Spin wave is a collective oscillation of spins around the direction of magnetization propagating in a wave-like manner in the ordering magnetic materials. There are some important properties of the spin waves to be used in logic devices. The minimum size of spin wave buses is limited by the wavelength of the transmitted signal, allowing for scaling down to the nanometer range. The typical group velocity of spin waves is of the order of $10^6$cm/s [6]. The coherence length of spin waves in ferromagnetic materials (e.g. NiFe) exceeds tens of microns at room temperature [6, 7], which allows us to utilize spin wave interference. Room temperature prototype devices based on interference effect have been recently demonstrated [8, 9]. We have previously considered different approaches to magnonic logic circuit construction e.g. dynamical circuits based on built-in spin wave amplifiers [10], and spin wave buses integrated with electronic non-linear elements [11]. In this work, we propose a combination of spin wave buses with multiferroic elements for building non-volatile logic circuits.

The rest of the paper is organized as follows. In the next section, we describe the principle of operation of magnonic circuits and the material structure of magneto-electric (ME) cell. We present the results of numerical simulations illustrating different modes of ME cell switching by spin waves in Section III. The examples of logic gates exploiting one or another switching mode are presented in Section IV. Discussions and Conclusions are given in sections V and VI, respectively.

**Section II. Principle of Operation**

The general view of a magnonic logic circuit is schematically shown in Fig.1. The circuit consists of magneto-electric (ME) cells connected by the spin wave buses. The cells are arranged in the columns. Input and the output ME cells have individual contacts, while the other cells in the circuit share the common column electrodes. ME cell is an synthetic multiferroic element possessing magnetic and electric polarizations. We consider ME cell as a bi-stable magnetic element, which magnetization states are used for information storage and controlled by the applied electric field. The cells are integrated with the spin wave buses and communicate via spin waves propagating through the spin wave buses. The functionality of magnonic logic circuit is defined by the number of ME cells and configuration of the spin wave buses.

The principle of operation of the magnonic circuit is as follows. Input data are received in the form of voltage pulses, which are applied to the input ME cells (e.g. +10mV corresponds to logic state 0, and −10mV corresponds to logic 1). The applied voltage affects the magnetic polarization of the ME cells (e.g. +10mV input produces clockwise magnetization rotation, and −10mV produces anticlockwise magnetization rotation). The change of magnetization of the ME cell results in the spin wave excitation in the spin wave buses. The initial phase of the spin wave signal is defined by the direction of magnetization rotation in the ME cell. Thus, the excited spin waves may have 0 or π phase difference depending the polarity of the applied voltage. Next, spin waves excited by the input ME cells propagate through the spin wave buses. At the point of intersection of two or several waveguides, spin waves interfere in constructive or destructive manner depending on the relative phase difference. The result of interference defines the magnetic polarization of the recipient ME cell. The switching of the ME cells is accomplished in a sequential manner column by column controlled by the global clock via the bias voltage applied to the column electrodes. Finally, spin wave signals reach the last column with the output ME cells. The read-out procedure is accomplished by detecting the induced voltage across the ME cell (e.g. +10mV corresponds to logic state 0, and −10mV corresponds to logic 1).

The material structure of the ME cell in conjunction with the spin wave bus is shown in Fig.2(A). The cell has a sandwich –like structure comprising from the bottom to the top a layer of a conducting magnetostrictive material (e.g. Ni, CoFe), a layer of piezoelectric (e.g. PZT –$PbZrTiO_3$), and a metallic contact (e.g. Al). The conducting ferromagnetic layer and the top electrodes serve as two sides of the parallel plate capacitor filled with the piezoelectric. An electric field applied across the piezoelectric layer produces stress. In turn, the stress affects the magnetostrictive material resulting in the change of the magnetization, and vice versa, a magnetic field applied to the magnetostrictive material produces stress, which affects the piezoelectric material changing its electric polarization. Such a cell comprising piezoelectric and ferromagnetic materials represents a synthetic two-phase multiferroic element [12] allowing magnetization control by the applied electric field. The ME cells are integrated within the spin wave buses in such a way, that a part of the spin wave bus is replaced by the magnetostrictive material of the ME cell. The latter insures the coupling between the magnetic moments of the magnetostrictive material of ME cell and the ferromagnetic material of the spin wave bus.

We assume the easy axes of the magnetic material of the ME cell and the magnetic material of the spin wave bus to be oriented in the perpendicular directions (e.g. for ME cell material along the Y axis, and for spin wave bus material along the Z axis). In this case, the coupling between the magnetic moments of two materials leads to the canted magnetization states. These states are schematically shown in Fig.2(B). The local magnetization of the ME cell may have two possible orientations $\pm M_y$ (along or opposite the Y axis), which in-plane component is much smaller than the out-of-plane $M_z$ component, $|M_y| \ll M_z$. The canted magnetization states are needed to implement the spin wave switching. It should be noted, that a propagating spin wave produces only a small deviation from the equilibrium magnetization orientation $\Delta M/M_s \ll 1$. Thus, spin wave itself cannot reverse (switch) the magnetization of the material it propagates through. However, it may be possible to switch a local magnetization between two polarization states if these states have a relatively small $\Delta M$ magnetization difference comparable with the spin wave amplitude. In the next Section, we present the results of numerical simulations showing the possibility of such switching.

The coexistence of the magnetic and electric polarizations gives us an intrigue possibility to exploit ME cells for spin wave excitation and detection. The input ME cells are aimed to convert electric pulses into the spin waves. An electric pulse applied to the input cell disturbs its magnetic polarization of the magnetostrictive material. The latter may be interpreted as an oscillation of the easy-axis direction caused by electro-mechanical-magnetic coupling. In turn, the oscillation of the anisotropy field will induce a local magnetization oscillation resulting in spin wave excitation. According to theoretical estimates [10], the use of magneto-electric coupling may provide an efficient mechanism for electric-to magnetic energy transfer. The same mechanism but in reverse order can be used for spin wave detection. A spin wave propagating through the magnetostrictive material induces stress, which affects the electric polarization of ME cell. A more detailed description of the ME cell operation is given in Ref. [5] and is not reproduced here.

A complete model for magnonic logic circuits simulation should take into consideration a number of physical processes including spin wave excitation by ME cells, spin wave propagation through the spin wave buses, ME cell switching, and spin wave to-voltage conversion. Some of these processes have been theoretically studied in our preceding theoretical [10, 13] and experimental [14, 15] works. In the present work, we restrict our consideration by modeling ME magnetization switching by spin waves. We consider a single ME cell integrated with a spin wave bus as shown in Fig.2(A). The objective of this study is to show the amplitude and phase dependent switching, which may lead to novel logic circuitry.

**III. Numerical modeling**

In order to illustrate the magnetization switching of the ME cell due tot the interaction with a spin wave, we carry out numerical simulations using the Landau-Lifshitz equation [16, 17]:

$$\frac{d\vec{m}}{dt} = -\frac{\gamma}{1+\alpha^2} \vec{m} \times \left[\vec{H}_{eff} + \alpha \vec{m} \times \vec{H}_{eff}\right], \tag{1}$$

where $\vec{m} = \vec{M}/M_s$ is the unit magnetization vector, $M_s$ is the saturation magnetization, $\gamma$ is the gyro-magnetic ratio, and $\alpha$ is the phenomenological Gilbert damping coefficient. The first term of equation (1) describes the precession of magnetization about the effective field and the second term describes its relaxation towards the direction of the field. The effective field is given as follows:

$$\vec{H}_{eff} = \vec{H}_d + \vec{H}_{ex} + \vec{H}_a + \vec{H}_b, \tag{2}$$

where $\vec{H}_d$ is the magnetostatic field ($\vec{H}_d = -\nabla \Phi$, $\nabla^2 \Phi = 4\pi M_s \nabla \cdot \vec{m}$), $\vec{H}_{ex}$ is the exchange field ($\vec{H}_{ex} = (2A/M_S)\nabla^2 \vec{m}$, $A$ is the exchange constant), $\vec{H}_a$ is the anisotropy field ($\vec{H}_a = (2K/M_S)(\vec{m} \cdot \vec{c})\vec{c}$, where $K$ is the uniaxial anisotropy constant, and $\vec{c}$ is the unit vector along the uniaxial direction $c \equiv (c_x, c_y, c_z)$), $\vec{H}_b$ is the external bias magnetic field. To simulate spin wave transport through the junction of two materials (ferromagnetic material of spin wave bus and magnetostrictive material of ME cell), one needs to now the material constants $M_s$, $A$, and $K$. for each material. For simplicity, we assumed all these parameters to be the same for both materials, expect the direction of the anisotropy field $\vec{c}$. We assume the material of the spin wave bus to have anisotropy field along the Z axis, and the material of the ME cell to have anisotropy field along the Y axis as shown in Fig.2. At the point of spin wave guide junction with ME cell, the interplay between the in-plane and out of plane magnetic field components leads to the canted magnetization states. We intentionally adjusted the material parameters ($M_s$, $A$, and $K$) to have the out of plane component of the effective magnetic field of the ME cell much higher than the in-plane component $H_\perp >> H_=$, so that there are two stable states for the magnetic material of ME cell (along to the Y axis and opposite to the Y axis), where $M_y << M_s$. In Fig. 3, we present the results of numerical modeling using Eq.(1) showing an example of such two states having $M_y$ projections ± 0.025$M_s$. Fig.3 shows the distribution of $M_y$ component along the X axis. The structure comprising ME cell and spin wave bus is modeled as a chain of 20 micro-magnetic cells. The 10$^{th}$ cell represents the ME cell. The easy axis of the 10$^{th}$ cell is along the Y axis, while the easy axis for all other cells (spin wave bus) is along the Z direction. The rest of numerical simulations are aimed to show the possibility of switching between the +0.025$M_s$ and - 0.025$M_s$ states as a result of interaction with a spin wave signal.

The spin wave signal is approximated by the wave packet equation for the magnetization components as follows [6]:

$$M_x = M_0 \times \exp(-t/\tau) \times \cos(k_0 x - \omega t + \phi),$$
$$M_y = M_0 \times \exp(-t/\tau) \times \sin(k_0 x - \omega t + \phi), \qquad (3)$$
$$M_z = \sqrt{M_s^2 - M_x^2 - M_y^2},$$

where $M_0$ is the spin wave amplitude, $\tau$ is the decay time, $x$ is the distance along the X-axis, $\omega$ is the spin wave frequency, $\phi$ is the initial phase.

First, we carry out numerical simulation on spin wave transport through the ME cell and studied the evolution of the ME cell magnetization as a function of the spin wave amplitude $M_0$. In Fig.4, we present three plots showing ME cell's in-plane magnetization trajectory simulated for three cases $M_0 \ll \Delta M$, $M_0 \gtrsim \Delta M$, and $M_0 \gg \Delta M$, where $\Delta M$ is the difference between the $M_y$ components for two stable states $(\Delta M = 0.05 M_s)$. The magnetization of the ME cell perform small oscillations around the steady state position ($M_y = +0.025 M_s$ or $M_y = -0.025 M_s$) if the spin wave amplitude $M_0$ is much less than $\Delta M$. No switching of magnetization occurs in this case as shown in Fig.4(A). The switching may occur if the spin wave amplitude is high enough to rotate the in-plane magnetization between the two steady states $M_0 \gtrsim \Delta M$, as illustrated in Fig.4(B). Hereafter, we define switching as the process of magnetization evolution where the final steady state is different from the initial state. In the ultimate limit of $M_0 \gg \Delta M$, the magnetization of the ME cell performs multiple rounds of in-plane rotation before it relaxes to the low-energy equilibrium state as shown in Fig.4(C). The final state may have $M_y = +0.025 M_s$ or $M_y = -0.025 M_s$, which depends on the shape of the spin wave packet. In Fig.5, we summarized the results of numerical simulations on ME cell magnetization switching as a function of the incoming spin wave amplitude. As one can see, the switching (magnetization evolution from one steady state to another) has a non-monotonic behavior. There is a threshold value for the spin wave amplitude $M_0 \approx \Delta M$, which allows for ME cell switching. High spin wave amplitude does not guarantee that the final magnetization state will be different from the initial state. As the amplitude of spin wave increases, the final state of the ME cell becomes dependent on the phase of the incoming spin wave signal.

In Fig.6, we present the results of numerical modeling showing spin wave switching as a function of the phase $\phi$ of the incoming spin wave packet ($M_0 = 0.1 M_s$). In this case, the amplitude of the incoming spin wave is higher than the barrier separating two steady states of magnetization. It turns out, that the final magnetization state coincides with the initial state if the phase of the incoming wave is within the one of the following regions $0<\phi<0.8\pi$ or $1.8<\phi<2\pi$. The cell changes its magnetization if the phase of the incoming spin wave is in the region $0.8<\phi<1.8\pi$. The final magnetization state depends also on the initial cell magnetization. In Fig.6(B), we show the results of numerical simulations for two initial cell states $+ 0.025 M_s$ and $M_y = -0.025 M_s$, respectively. There are equal phase regions leading to switching form positive to negative $M_y$ projections, and vise versa.

To summarize this section, we observe three different modes for ME cell switching dynamics. First, no switching occurs if the amplitude of the incoming spin wave is not big enough to overcome the potential barrier between the two canted magnetization states $M_0 << \Delta M$. In this case, a weak spin wave produces magnetization oscillation around the canted state. As the amplitude of the spin wave increases, the switching may take place. In this mode $M_0 \gtrsim \Delta M$, the final state depends on two factors: the phase of the incoming spin wave, and the initial state of the recipient ME cell. In the ultimate limit $M_0 >> \Delta M$, the final state of ME cell depends mostly on the structure of the incoming spin wave packet (e.g. phase $\phi$, packet width $2/\delta$, damping parameter $\alpha$).

**Section IV    Magnonic logic circuits**

Computation within the magnonic circuit is associated with change of ME cell magnetization as a function of the magnetization of other cells. Translating magnetization states to the logic state, we define logic 0 and 1 as two magnetization sates for $M_y$ (e.g. corresponding to + $0.025M_s$ and $-0.025M_s$, respectively). The results of numerical simulations show the modes of ME cell switching, where the final ME magnetization depends on the phase of the incoming spin wave. The phase of the incoming wave is defined by the magnetization of the wave emitting ME cell as well as on the phase change accumulated during the propagation though the spin wave bus. The latter makes it possible to engineer magnonic logic circuits by exploiting spin wave buses as passive elements for phase modulation.

To illustrate this idea, we show in Fig.7 the examples of circuits consisting of the same number of ME cells, which functionality is controlled by the distance between the input and output cells. For simplicity, we introduce two characteristic lengths for the connecting spin wave buses: $l_0$ and $l_\pi$, which provide 0 and $\pi$ phase shifts, respectively. Approximately, $l_0$ and $l_\pi$ can be estimated as follows $l_0=n\times\lambda$, and $l_\pi=(n+1/2)\times\lambda$, where $n=1,2,3,...$ , and $\lambda$ is the wavelength, $\lambda=2\pi/k$. Two cells separated by distance $l_0$ can operate as a Buffer logic gate (Fig.7(a)). The same two cells separated by distance $l_\pi$ providing perform a NOT logic gate (Fig.7(B)). In other to build the Buffer and NOT gates, the switching should occur at high amplitude mode $M_0 >> \Delta M$, where the final state of the output state does not depend on the initial state.

The same approach can be applied to the two-input and one-output gates. In the case of two-input gates, two spin waves coming to the output cell may have a π relative phase difference and interfere destructively providing a net zero magnetization change to the recipient ME cell. In this case, the final ME cell will be defined by its initial state. In Fig.7(C-D), there are shown circuits consisting of two input and one output cell. The circuit operates as AND gate if the input and the output cells are separated by distance $l_0$, and the output cell is set up to state 0 before the computation. The similar three-cell circuit can operate as NAND gate if the distance between the input and the output cells is $l_\pi$, and the output cell is set up to logic state 1 prior to computation. It is also possible to build two-input and one-output logic gates by using one of the cells as the input and the output cell. In this case, the final state of the output cell depends on the phase of the incoming spin wave and its initial state.

Next, we show the example of three-input one-output MAJ gates in Fig.8. The input cells and the output cell are separated by the distance $l_0$. Three input waves may have 8 possible combinations of the input phases $(0,0,0)$, $(0,0,\pi)$, $(0,\pi,0)$,... $(\pi,\pi,\pi)$. As a result of spin wave interference, the resultant magnetization is defined by the majority of the input phases. In this scenario, the amplitude of the input spin waves has to be much higher than the barrier between the ME state to ensure the initial state independent switching $M_0 >> \Delta M$. Thus, the output cell will have final magnetization corresponding to the majority of the inputs. MAJ logic gate is a powerful element for logic construction. In general, Majority logic is more powerful for implementing a given digital function with a smaller number of logic gates than CMOS [18].

Evolving the idea of using spin wave buses as passive logic elements for phase modulation, we consider the possibility of making magnetic waveguides of different width, or/and composition to provide the desired phase shift at the same propagation time. For example, an addition a layer of ferromagnetic material (e.g. NiFe) on the top of wave bus would provide an additional phase shift to the propagating spin waves due to the dipole-dipole interaction. In principle, it is possible to design spin wave buses of the different length providing different phase shift at the same propagation time. The latter let us construct more sophisticated logic gates such as the XOR gate shown in Fig.9. To engineer this gate, it is required to provide a relative $\pi$ phase shift for the two input spin waves. In this case, the waves interfere constructively if they have different initial phases and destructively if they have the same initial phase. The output cell should be set up to the logic state 0 before the computation. The input waves interfere destructively and does not change the output state, while two wave interfering constructively change the state of the output cell. An essential condition for the XOR gate operation is that the input spin waves have to arrive to the output cell at the same time, which explains the need in the special waveguides providing 0 or $\pi$ phase shifts but for the same propagation time.

In the considered above magnonic circuits, the input spin waves have to arrive to the output cell at the same time. It is interesting to note the possibility of building logic gates utilizing sequential switching (e.g. spin waves excited by the input cells arrive to the output cell one after another). An example of the sequential logic gate is shown in Fig.10. There are three input cells and one output cell. It takes longer time for spin wave excited by Input-3 and Input-2 to travel to the output then for the wave excited by Input-1. The initial state of the output cell is set up to 0 prior to computation. The operation of this gate is as follows. The excitation pulses are applied to all three input cells at the same time. First, the output cell receives the spin wave from Input-1. The first switching is a simple Buffer gate as shown in Fig.7(A). The output cell changes its state to1, if and only if, the input is 1. Next, two waves form Input-2 and Input-3 arrive at the same time. The second switching is similar to the XOR gate operation. The waves coming form Input-2 and Input-3 accumulate an additional $\pi$-phase shift during the propagation. The waves interfere constructively and change the state of the output cell if and only if the logic states of the inputs 2 and 3 are different (e.g. 0,1 or 1,0). The waves interfere destructively if the logic inputs are the same (e.g. 0,0 or 1,1). The truth table is shown in the insert to Fig. 10. The operation of this logic gate resembles a modular function providing a mod2 output to the sum of the inputs (e.g. $0+0+0\equiv0mod2$, $0+0+1\equiv1mod2$, $0+1+1\equiv0mod2$, $1+1+1\equiv1mod2$).

The presented examples show different ways of building logic gates taking advantage of phase modulation in the connecting spin wave buses. There is a plethora of possible logic circuits, which can be constructed by utilizing one or another ME switching modes or combination of simultaneous or sequential switching. The most important advantage of the considered wave-based circuits is the ability to perform logic operations in wires, by changing the phase of the propagating signal. The latter provides a powerful tool for building logic circuits with a fewer number of elements than required for CMOS technology. In Fig.11, we show the design of the magnonic Full Adder circuit. The Full Adder circuit adds three one-bit binary numbers (A, B, and $C_0$-input carry) and outputs two one-bit binary numbers, a sum (S) and a carry ($C_1$). The truth table is shown in Fig.11. The $C_1$ output is nothing but the MAJ gate for the three inputs (A,B,$C_0$). And the S output is the MOD2 gate for the same inputs. The presented design in Fig.12 shows the most compact circuit structure with minimum possible number of elements. There are just 5 ME cells (three input cells, and two output cells) connected with spin wave buses providing different phase shifts to the propagating spin waves. In contrast, a transistor-based implementation requires a larger circuit with seven or eight gate elements (about 25–30 MOSFETs) [19].

V. Discussion

In order to find practical application, magnonic logic circuits have to show capabilities beyond the conventional CMOS-based logic circuits in terms of functional throughput and/or lower power consumption. The principle of operation of magnonic circuits is different from the conventional CMOS technology, as there are no magnetic transistors (e.g. devices transmitting or stopping spin waves as a function of the external field). The comparison between the magnonic and CMOS-based logic devices should be done at the circuit level by comparing the overall circuit parameters such the number of functions per area per time, time delay per operation, and energy required for logic function. In this Section, we present the estimates on the magnonic circuits and compare them with the CMOS circuits.

*Scalability* of magnonic logic circuit is defined by several parameters: the size of ME cell ($F \times F$); the number of ME cells per circuit $N_{ME}$; the length $L_{swb}$ and the width $W_{swb}$ of the spin wave buses. These parameters are related to each other via the same physical quantity - the wavelength of the spin wave $\lambda$. As we mentioned in the previous Section, the length of the spin wave buses providing 0 or $\pi$ phase shift to the propagating spin wave is defined by the wavelength. For example, the minimum length of the Inverter gate cannot be shorter than $\lambda/2$. Theoretically, the feature size $F$ of the ME cell can be much smaller than the wavelength $\lambda$ of the information carrying spin waves. On the other hand, the length of the ME cells should be about the wavelength $F \sim \lambda$ for efficient spin wave excitation via the magneto-electric coupling. The width of the spin wave bus $W_{swb}$ is also related to the wavelength $\lambda$ via the dispersion law. However, the width of the spin wave bus can be much smaller than the wavelength. In our estimates, we assume the feature size of the ME cell $F$ to be equal the wavelength $\lambda$ ($F \approx \lambda$), $W_{swb} \ll L_{swb}$, and $L_{swb}$ to be one or one and a half of the wavelength depending the particular logic circuit (e.g. $L_{swb}=\lambda$ for Buffer gate, $L_{swb}=\lambda/2$ for Inverter). The number of ME cell per circuit varies depending circuit functionality. At this moment, there is no empirical rule to estimate the size of the magnonic logic circuits based on the number of ME cells. Below, we present estimates for the area $A$ of some logic circuits.

$A = F \times (2F+\lambda) \approx 2\lambda^2$           - Buffer

$A = F \times (2F+\lambda/2) \approx 2.5\lambda^2$       - Inverter

$A = F \times (3F + \lambda + \lambda) \approx 3\lambda^2$      - AND gate

$A = (3F + 2\lambda) \times (2F + \lambda) \approx 15\lambda^2$    - MAJ gate/MOD2 gate

$A = (3F + 2\lambda) \times (3F + 2\lambda) \approx 25\lambda^2$    - Full Adder Circuit.

*Time delay* per circuit is a sum of the following: the time required to excite spin waves by the input ME cells $t_{ext}$, propagation time for spin waves from the input to the output cells $t_{prop}$, and the time of magnetization relaxation in the output ME cells $t_{relax}$:

$t_{delay} = t_{ext} + t_{prop} + t_{relax}$

The propagation time can be estimated by dividing the length of the spin wave bus connecting the most distant input and the output cells by the spin wave group velocity $v_g$, $t_{prop} = L_{swb}/v_g$. The group velocity depends on the material and geometry of the bus as well as the specific spin wave mode. The typical group velocity for magnetostatic spin waves propagating in conducting ferromagnetic materials (e.g. NiFe) is about $10^6$ cm/s [6, 7]. The relaxation time of the output ME cell depends on the material properties of the magnetostrictive material (e.g. damping parameter $\alpha$) and can be estimated by micromagnetic simulations. More difficult is to estimate the time required for spin wave excitation $t_{ext}$. The time required to excite spin waves by ME cells depends on the number of parameters (e.g. strength of ME coupling, material properties of the magnetostrictive material, size of the ME cell). At any rate, the minimum time delay for spin wave excitation is limited by the *RC* delay of the electric part, where *R* is the resistance of metallic interconnects, and *C* is the capacitance of the ME cell. The lack of experimental data makes it difficult to estimate the practically achievable excitation time. As a rough estimate, we assume the total delay per circuit to be defined by the spin wave propagation time:

$$t_{delay} \approx L_{swb}/v_g$$

We want to note, that the time delay for magnonic logic circuits does not increase with the increase of the number of ME cells. All circuits shown in Fig.5-7, would have approximately the same delay. There will be a time difference of $\Delta t=0.5\lambda/v_g$, for the circuits having spin wave buses of the length $l_0$ and $l_\pi$.

*Power consumption* is another critical parameter to be estimated. We want to emphasize that the described magnonic logic circuits are inherently non-volatile. Being switched, ME cells can preserve the result of computation for long time, which depends on the energy barrier for the two magnetization states. To be practical, the energy barrier should be about $40kT$, where $k$ is the Boltzmann constant, and $T$ is the ambient temperature. Magnonic circuits consume energy only during the computation, and no outer power source is required to preserve the results of computation. The overall energy consumed by the magnonic logic $E$ is defined by the number of the ME cells and the energy required for spin wave excitation $E_{ME}$:

$$E = N_{ME} \times E_{ME}.$$

The energy of spin wave must be high enough to exceed the potential barrier separating the two canted states of the recipient ME cell. The latter defines the minimum energy of the spin waves. The energy per excitation should include not only the spin wave energy but also account for the losses inside the ME cell. In general, the fundamental limit for the conversion efficiency of ME cell made of a parallel-plate capacitor filled by piezoelectric-magnetostrictive composites is defined by the ratio between the magnetic and electro-mechanical losses. Theoretically, the conversion efficiency of the two-phase multiferroic structure can be high (up to 97%) for the ME structure precisely matching electro-mechanical and magnetic resonances [10] [20]. For example, assuming 10% conversion efficiency for a single ME cell, the total energy per function for the Full Adder Circuit can be estimated as follows:

$E_{Adder} \approx 5 \times 10 \times 40 kT = 24$ aJ.

In Table I, we summarized the estimates for magnonic Full Adder circuit and compare them with the parameters of the CMOS-based circuit. The data for the Full Adder circuit made on 45nm and 32nm CMOS technology is based on the ITRS projections [21] and available data on current technology [22]. The data for the magnonic circuits is based on the design shown in Fig.11 and the above made estimates. Magnonic logic circuits may have significant advantage in minimizing circuit area due to the fewer number of elements required per circuit (e.g. 5 ME cells versus 25-30 CMOSs). We estimated the time delay for the magnonic circuit as the time required for spin wave to propagate from the input to the output cell, which is the shortest possible delay time. Even in this best scenario, magnonic logic circuits would be slower than the CMOS counterparts. Nevertheless, the overall functional throughput may be higher for magnonic logic circuits due to the smaller circuit area. The most prominent advantage over CMOS circuitry is expected in minimizing power consumption. There is no static power consumption in magnonic logic circuits based on non-volatile magnetic cells. The utilization of multiferroic materials together with low-energy spin wave switching may substantially reduce the energy per operation.

The optimistic projections on magnonic logic circuit efficiency are based on many assumptions and theoretical predictions. There is a number of problems to be solved before magnonic logic devices will be able to compete with CMOS-based circuits. The most critical concerns are associated with the feasibility of ME cell multi-functional operation, and system robustness (e.g. defect and size variability tolerance). To the best of our knowledge, there is no prototype ME cell, which can perform all of the desired functions (e.g. spin wave excitation, information storage, switching by spin wave, and spin wave to voltage conversion). Experimentally obtained multiferroic structures comprising piezoelectric and magnetostrictive materials show prominent magneto-electric coupling PZT/NiFe$_2$O$_4$ (1,400 mV cm$^{-1}$ Oe$^{-1}$)[23], CoFe$_2$O$_4$/BaTiO$_3$ (50 mV cm$^{-1}$ Oe$^{-1}$)[24], PZT/Terfenol-D (4,800 mV cm$^{-1}$ Oe$^{-1}$)[25]. Though, it should be noted that the experiment values of the magneto-electric coupling are obtained for DC applied electric field. It is not clear if the composite structures can sustain high-frequency frequency operation. There are other approaches to magneto-electric cell construction (e.g. voltage-controlled surface anisotropy [26]), which may be more efficient than the combination of piezoelectric-magnetostrictive materials.

Size variability will be the main factor limiting magnonic logic circuits scalability. The operation of the circuits is based on the phase-dependent switching. The variation of the spin wave phases will be critical for circuit operation. For example, the permissible length variation of the one-input one-output spin wave buses can be estimated as $\lambda/8$, which gives a $\pi/4$ phase change. The required accuracy increases with the increase of the number of interfering spin waves. ME cell size variation is be even more important, as the small variation of the cell geometry/composition may significantly affect the shape of the emitted spin wave signal. Global clock operation and time synchronization required for magnonic circuits are the other issues, which require special consideration.

Nevertheless these critical comments, the proposed concept of magnonic logic circuit engineering shows a fundamental advantage of using phases in addition to amplitudes for minimizing the number of elements per logic gate. At some point, the evolution of magnetization via interaction with waves resembles the operation of a quantum computer [27]. It is an interesting question to ask: Is it feasible to build a classical wave-based computing device with functional capabilities close to the quantum computers? It has been shown that it is possible to exploit classical wave interference and superposition techniques to implement algorithms where quantum entanglement is not required (e.g. Deutsch-Jozsa algorithm [28], Grover Search algorithm [29], Bernstein-Vazirani algorithm [30]). The latter gives an intrigue possibility to build a new class of wave-based logic devices with capabilities intermediate between the conventional transistor-based and purely quantum computers. The advantage of using waves for information transmission and processing would ever increase with the increase of the number of processing bits. It is not mandatory for magnonic logic to outperform CMOS in all possible logic gates. It would be of great practical benefit to build special class of logic devices able to complement CMOS for special task data processing (e.g. image processing, speech recognition). Magnonic logic circuits have a great potential to fulfill this task.

## VI. Conclusions

We have described a concept of magnetic logic devices taking advantage of phase-dependent switching. The latter makes it possible to exploit magnetic waveguides as passive logic elements for phase modulations. We presented a library of logic gates comprising magneto-electric cells and spin wave buses providing 0 or π phase shifts. The utilization of waveguides for phase modulation allows us to construct sophisticated logic gates such as MOD2 gate, and Full Adder circuit with minimum number of elements. Nonvolatility is another important property of the proposed magnetic circuits allowing to eliminate the cause of the static power consumption. According to the estimates, magnonic logic circuits may provide higher functional throughput for less consumed energy than the CMOS circuits. There is a number of questions regarding the operation of the magneto-electric cells and overall system stability with respect to the size variation, which require additional study. In summary, the proposed concept of magnetic logic circuit engineering offers a fundamental advantage over the CMOS technology and may provide a route to a wave-based logic circuitry with capabilities far beyond the limits of the traditional transistor-based approach.

**Acknowledgments**

The work was supported by the DARPA program on Non-volatile Logic (program manager Dr. Devanand K. Shenoy) and by the Nanoelectronics Research Initiative (NRI) (Dr. Jeffrey J. Welser, NRI Director) via the Western Institute of Nanoelectronics (WIN).
25

References:

[1] G. E. Moore, "Gramming more components onto integrated circuits," *Electronics*, vol. 38, pp. 114-117, 1965.
[2] A. K. Geim, "Graphene: Status and Prospects," *Science*, vol. 324, pp. 1530-34, 2009.
[3] A. Khitun and K. Wang, "Nano scale computational architectures with Spin Wave Bus," *Superlattices & Microstructures*, vol. 38, pp. 184-200, 2005.
[4] A. Khitun, M. Bao, and K. L. Wang, "Spin Wave Magnetic NanoFabric: A New Approach to Spin-based Logic Circuitry," *IEEE Transactions on Magnetics*, vol. 44, pp. 2141-53, 2008.
[5] A. Khitun, M. Bao, and K. L. Wang, "Magnetic cellular nonlinear network with spin wave bus for image processing," *Superlattices & Microstructures*, vol. 47, pp. 464-83, 2010.
[6] M. Covington, T. M. Crawford, and G. J. Parker, "Time-resolved measurement of propagating spin waves in ferromagnetic thin films," *Physical Review Letters*, vol. 89, pp. 237202-1-4, 2002.
[7] T. J. Silva, C. S. Lee, T. M. Crawford, and C. T. Rogers, "Inductive measurement of ultrafast magnetization dynamics in thin-film Permalloy," *Journal of Applied Physics*, vol. 85, pp. 7849-62, 1999.
[8] M. P. Kostylev, A. A. Serga, T. Schneider, B. Leven, and B. Hillebrands, "Spin-wave logical gates," *Applied Physics Letters*, vol. 87, pp. 153501-1-3, 2005.
[9] Y. Wu, M. Bao, A. Khitun, J.-Y. Kim, A. Hong, and K. L. Wang, "A Three-Terminal Spin-Wave Device for Logic Applications," *J. Nanoelectron. Optoelectron*, vol. 4, pp. 394–397, 2009.
[10] A. Khitun, D. E. Nikonov, and K. L. Wang, "Magnetoelectric spin wave amplifier for spin wave logic circuits," *Journal of Applied Physics*, vol. 106, pp. 123909-7, 2009.
[11] A. Khitun, M. Bao, J.-Y. Lee, K. L. Wang, D. W. Lee, S. X. Wang, and I. V. Roshchin, "Inductively Coupled Circuits with Spin Wave Bus for Information Processing," *Journal of Nanoelectronics and Optoelectronics*, vol. 3, pp. 24-34, 2008.
[12] W. Eerenstein, N. D. Mathur, and J. F. Scott, "Multiferroic and magnetoelectric materials," *Nature*, vol. 442, pp. 759-65, 2006.
[13] A. Khitun, D. E. Nikonov, M. Bao, K. Galatsis, and K. L. Wang, "Efficiency of spin-wave bus for information transmission," *IEEE Transactions on Electron Devices*, vol. 54, pp. 3418-21, 2007.
[14] A. Khitun, M. Bao, Y. Wu, J.-Y. Kim, A. Hong, A. Jacob, K. Galatsis, and K. L. Wang, "Logic Devices with Spin Wave Buses - an Approach to Scalable Magneto-Electric Circuitry," *Mater. Res. Soc. Symp. Proceedings*, vol. 1067, pp. B01-04, 2008.
[15] Y. Wu, M. Bao, A. Khitun, J.-Y. Kim, A. Hong, and K. L. Wang, "A Three-Terminal Spin-Wave Device for Logic Applications," *J. Nanoelectron. Optoelectron*, vol. 4, pp. 394–397, 2009.

[16] L. D. Landau and E. M. Lifshitz, "Theory of the dispersion of magnetic permeability in ferromagnetic bodies," *Phys. Z. Sowietunion*, vol. 8, pp. 153, 1935.
[17] T. L. Gilbert, *Physical Review*, vol. 100, pp. 1243, 1955.
[18] A. R. Meo, "Majority Gate Networks," *IEEE Transactions on Electronic Computers*, vol. EC-15, pp. 606-18, 1966.
[19] T. Oya, T. Asai, T. Fukui, and Y. Amemiya, "A majority-logic device using an irreversible single-electron box," *IEEE Transactions on Nanotechnology*, vol. 2, pp. 15-22, 2003.
[20] D. Shuxiang, L. Jie-Fang, and D. Viehland, "Magnetoelectric coupling, efficiency, and voltage gain effect in piezoelectric-piezomagnetic laminate composites," *Journal of Materials Science*, vol. 41, pp. 97-106, 2006.
[21] "International Technology Roadmap for Semiconductors," *Semiconductor Industry Association*, 2007.
[22] A. Chen, *private communication*, 2010.
[23] G. Srinivasan, E. T. Rasmussen, J. Gallegos, R. Srinivasan, I. Bokhan Yu, and V. M. Laletin, "Magnetoelectric bilayer and multilayer structures of magnetostrictive and piezoelectric oxides," *Physical Review B (Condensed Matter and Materials Physics)*, vol. 64, pp. 2144081-6, 2001.
[24] J. Van Den Boomgaard, D. R. Terrell, R. A. J. Born, and H. Giller, "An in situ grown eutectic magnetoelectric composite material. I. Composition and unidirectional solidification," *Journal of Materials Science*, vol. 9, pp. 1705-9, 1974.
[25] R. Jungho, V. Carazo, K. Uchino, and K. Hyoun-Ee, "Magnetoelectric properties in piezoelectric and magnetostrictive laminate composites," *Japanese Journal of Applied Physics, Part 1 (Regular Papers, Short Notes & Review Papers)*, vol. 40, pp. 4948-51, 2001.
[26] T. Maruyama, Y. Shiota, T. Nozaki, K. Ohta, N. Toda, M. Mizuguchi, A. A. Tulapurkar, T. Shinjo, M. Shiraishi, S. Mizukami, Y. Ando, and Y. Suzuki, "Large voltage-induced magnetic anisotropy change in a few atomic layers of iron," *Nature Nanotech*, vol. 4, pp. 158 - 161, 2009.
[27] A. Khitun, R. Ostroumov, and K. L. Wang, "Spin-wave utilization in a quantum computer," *Physical Review A*, vol. 64, pp. 062304/1-5, 2001.
[28] E. Brainis, L. P. Lamoureux, N. J. Cerf, P. Emplit, M. Haelterman, and S. Massar, "Fiber-optics implementation of the Deutsch-Jozsa and Bernstein-Vazirani quantum algorithms with three qubits," *Physical Review Letters*, vol. 90, pp. 157902/1-4, 2003.
[29] P. G. Kwiat, J. R. Mitchel, P. D. D. Schwindt, and A. G. White, "Grover's search algorithm: an optical approach," *Journal of Modern Optics*, vol. 47, pp. 257-66, 2000.
[30] Arvind, G. Kaur, and G. Narang, "Optical implementations, oracle equivalence, and the Bernstein-Vazirani algorithm," *Journal of the Optical Society of America B (Optical Physics)*, vol. 24, pp. 221-5, 2007.

Figure captions

Fig. 1. Schematic view of a magnonic logic circuit. Input data are received in the form of voltage pulses, which are applied to the input ME cells (e.g. +10mV corresponds to logic state 0, and −10mV corresponds to logic 1). ME cell is an artificial multiferroic element possessing magnetic and electric polarizations. The applied voltage affects the magnetic polarization of the ME cells (e.g. +10mV input produces clockwise magnetization rotation, and −10mV produces anticlockwise magnetization rotation). Input cells emit spin waves, which propagate through the spin wave buses. At the point of intersection of two or several buses, spin waves interfere in constructive or destructive manner depending on the relative phase difference. The result of interference defines the magnetic polarization of the recipient ME cell. The switching of the ME cells is accomplished in a sequential manner column by column controlled by the global clock. Finally, spin wave signals reach the last column with the output ME cells. The read-out procedure is accomplished by detecting the induced voltage across the ME cell (e.g. +10mV corresponds to logic state 0, and −10mV corresponds to logic 1).

Fig.2 (A). Material structure of ME cell in conjunction with spin wave bus. The cell consists from a layer of a conducting magnetostrictive material (e.g. Ni, CoFe), a layer of piezoelectric (e.g. PZT –$PbZrTiO_3$), and a metallic contact (e.g. Al) on the top. The ME cells are integrated within the spin wave buses in such a way, that a part of the spin wave bus is replaced by the magnetostrictive material of the ME cell. (B) Local magnetization

of the ME cell may have two possible orientations $\pm M_y$ (along or opposite the Y axis), $|M_y| \ll M_z$.

Fig.3. Results of numerical modeling showing the distribution of $M_y$ component along the chain of 20 magnetic cells. X-axis. The 10$^{th}$ cell represents the ME cell. The easy axis of the 10$^{th}$ cell is along the Y-axis, while the easy axis for all other cells (spin wave bus) is along the Z direction. There are two steady state magnetic configurations for the ME cell with $M_y$ projections $\pm 0.025 M_s$.

Fig.4. In-plane magnetization trajectory of the ME cell simulated for three cases (A) $M_0 \ll \Delta M$, (B) $M_0 \gtrsim \Delta M$, (C) $M_0 \gg \Delta M$, where $M_0$ is the spin wave amplitude, and $\Delta M$ is the difference between the $M_y$ components for two stable states *($\Delta M=0.05 M_s$)*. The magnetization of the ME cell perform small oscillations around the steady state position. No switching of magnetization occurs in case (A). The switching may occur if the spin wave amplitude is high enough to rotate the in-plane magnetization between the two steady states in case (B). The magnetization of the ME cell performs multiple rounds of in-plane rotation before it relaxes to the low-energy equilibrium state as shown in case (C).

Fig.5. Summary of numerical simulations showing the ME cell magnetization switching as a function of amplitude $M_0$ of the incoming spin wave *($\phi=0$, $\Delta M=0.05 M_s$)*. The switching is a non-monotonic function of the amplitude. There is a threshold value for the spin wave amplitude $M_0 \approx \Delta M$, which allows for ME cell switching. At higher spin

wave amplitudes, the final state of the ME cell dependents on the phase of the incoming spin wave.

Fig.6. (A) Summary of numerical simulations showing the ME cell magnetization switching from $M_y = + 0.025M_s$ to $M_y = - 0.025M_s$ as a function of the phase $\phi$ of incoming spin wave ($M_0 = 0.1M_s$, $\Delta M=0.05M_s$). (B) The blue and the red curves show switching from $+ 0.025M_s$ to $- 0.025M_s$ and from $- 0.025M_s$ to $+ 0.025M_s$, respectively ($M_0 = 0.1M_s$, $\Delta M=0.05M_s$).

Fig.7. Examples of magnonic logic gates which functionality is controlled by the phase delay (A) Buffer gate, (B) NOT gate, (C) AND gate, and (D) NAND gate. Spin wave buses of length $l_0$ provides a zero-phase shift, and a bus of length $l_\pi$ provides a $\pi$-phase shifts. The switching is at high amplitude mode $M_0>>\Delta M$ for the Buffer and NOT gates. The switching is in the $M_0 \gtrsim \Delta M$ mode for the AND and NAND gates. The output cell should be set up to logic state 0 for AND gate, and to logic state 1 for NAND gate prior to computation.

Fig.8. Schematics of the three-input one-output magnonic Majority (MAJ) gate. The output cell receives three waves form the input cells, which may have one of the eight possible phase combinations $(0,0,0)$, $(0,0,\pi)$, $(0,\pi,0)$,... $(\pi,\pi,\pi)$. The resultant magnetization of the output cell is defined by the result of spin wave interference (the majority of the input phases).

Fig.9. Schematics of the XOR gate. The spin wave buses connecting the input and the output cell are designed to provide a relative π-phase shift for the two input spin waves (e.g. one of the buses provide 0-phase shift, and the second one provides a π-phase shift). The input waves arrive to the output cell at the same time. The waves interfere constructively and change the output cell magnetization if and only if the inputs states are (0,1) or (1,0). The input waves interfere destructively and does not change the output state for (0,0) and (1,1) cases. The output cell should be set up to the logic state 0 before the computation.

Fig.10. Schematics of the MOD2 gate. It is an example of sequential switching. The output cell receives first the wave from Input-1 (Buffer gate). Next, the output cell receives two waves from Input-2 and Input-3. The second switching is similar to the XOR gate. The state of the output is changed if and only if the waves from inputs 2 and 3 have a relative π-phase difference (e.g. 0,1 or 1,0 states). The operation of this logic gate resembles a modular function providing a mod2 output to the sum of the inputs (e.g. 0+0+0≡0mod2, 0+0+1≡1mod2, 0+1+1≡0mod2, 1+1+1≡1mod2).

Fig.11. Schematics of the magnonic Full Adder circuit. It constructed as a combination of the MAJ and MOD2 gates. The whole circuit consists of just 5 ME cells (three input cells, and two output cells) connected with spin wave buses providing different phase shifts to the propagating spin waves.

Fig.12 Estimates and comparison between CMOS and magnonic Full Adder circuit. The data for CMOS-based circuit is based on the ITRS-2007 projections. The data for the magnonic circuits is for the design shown in Fig.11.

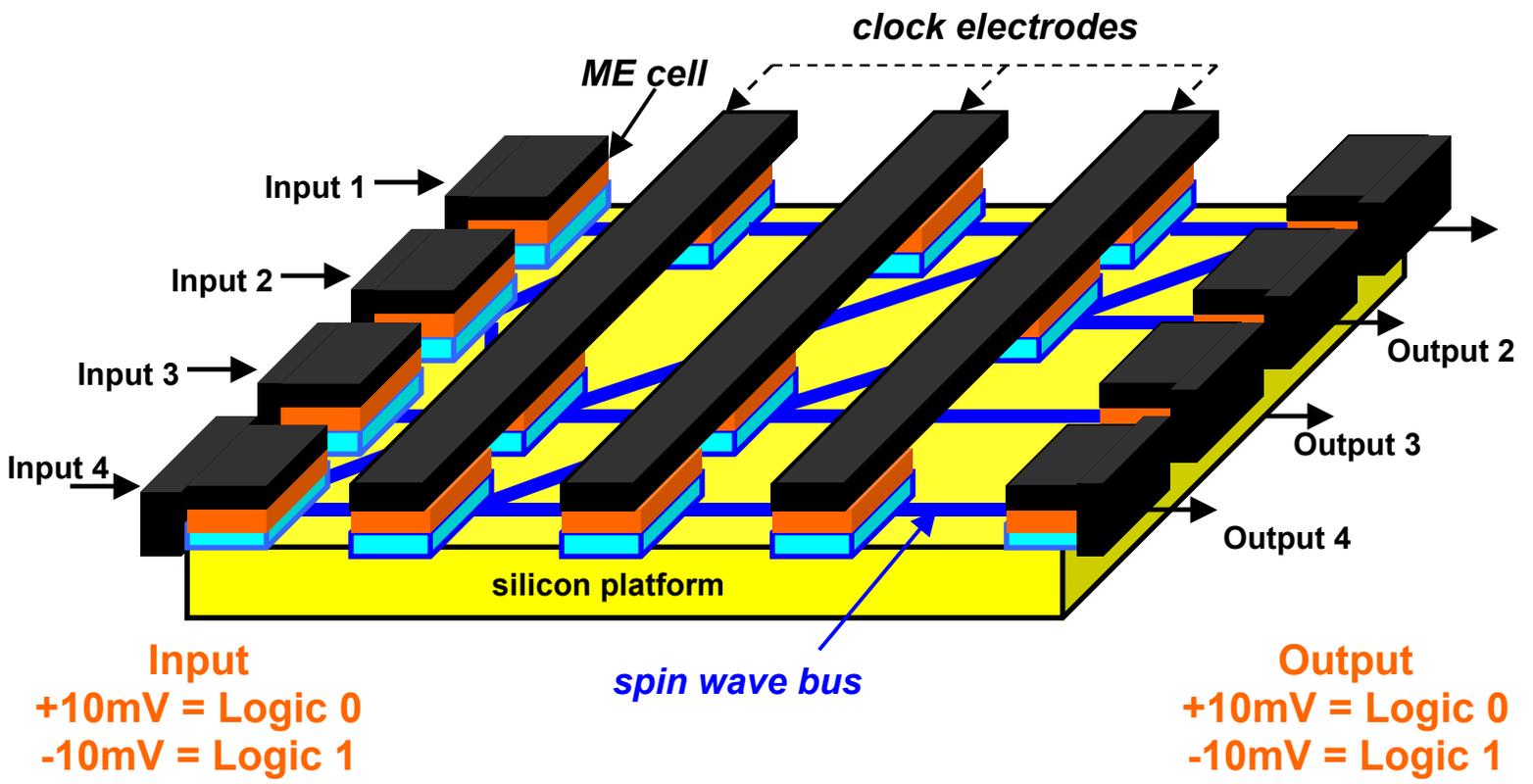

**Fig.1**

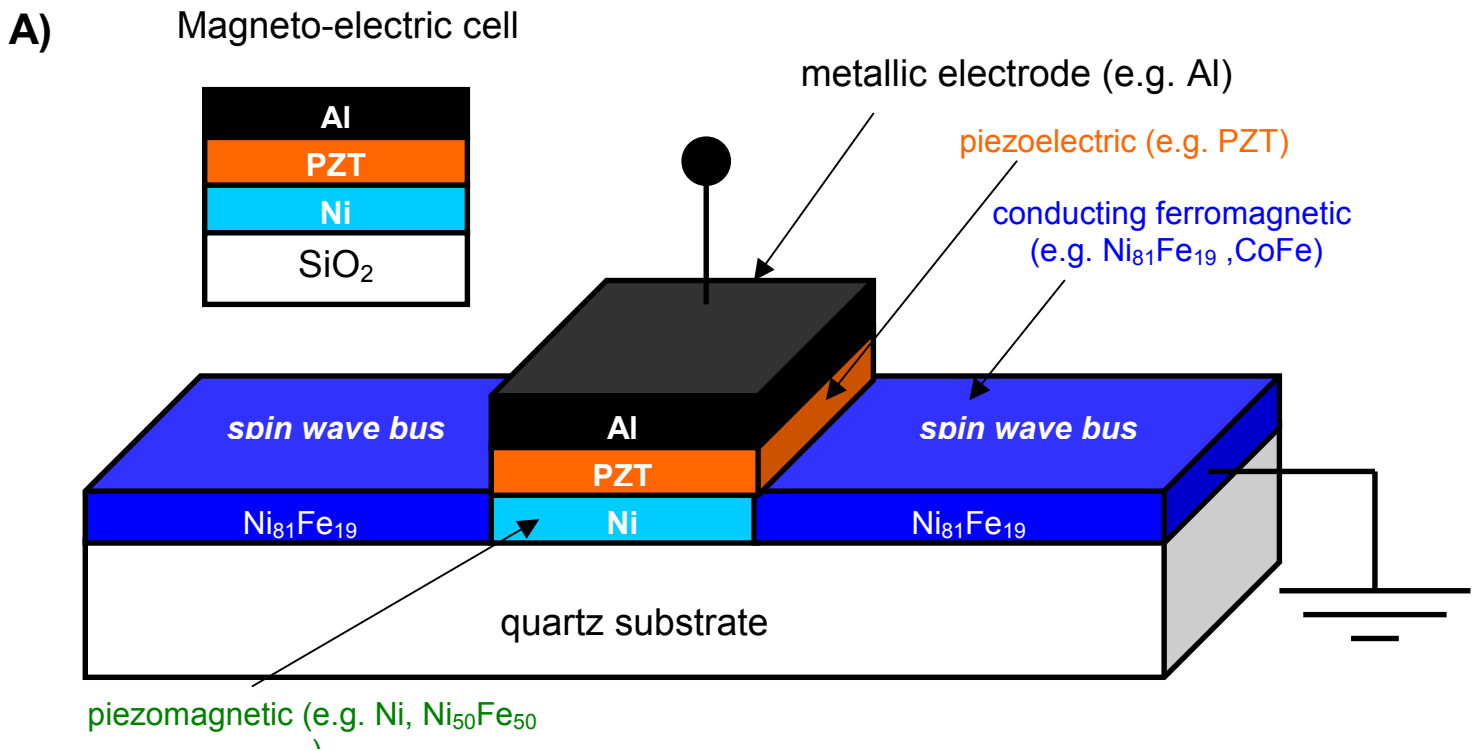

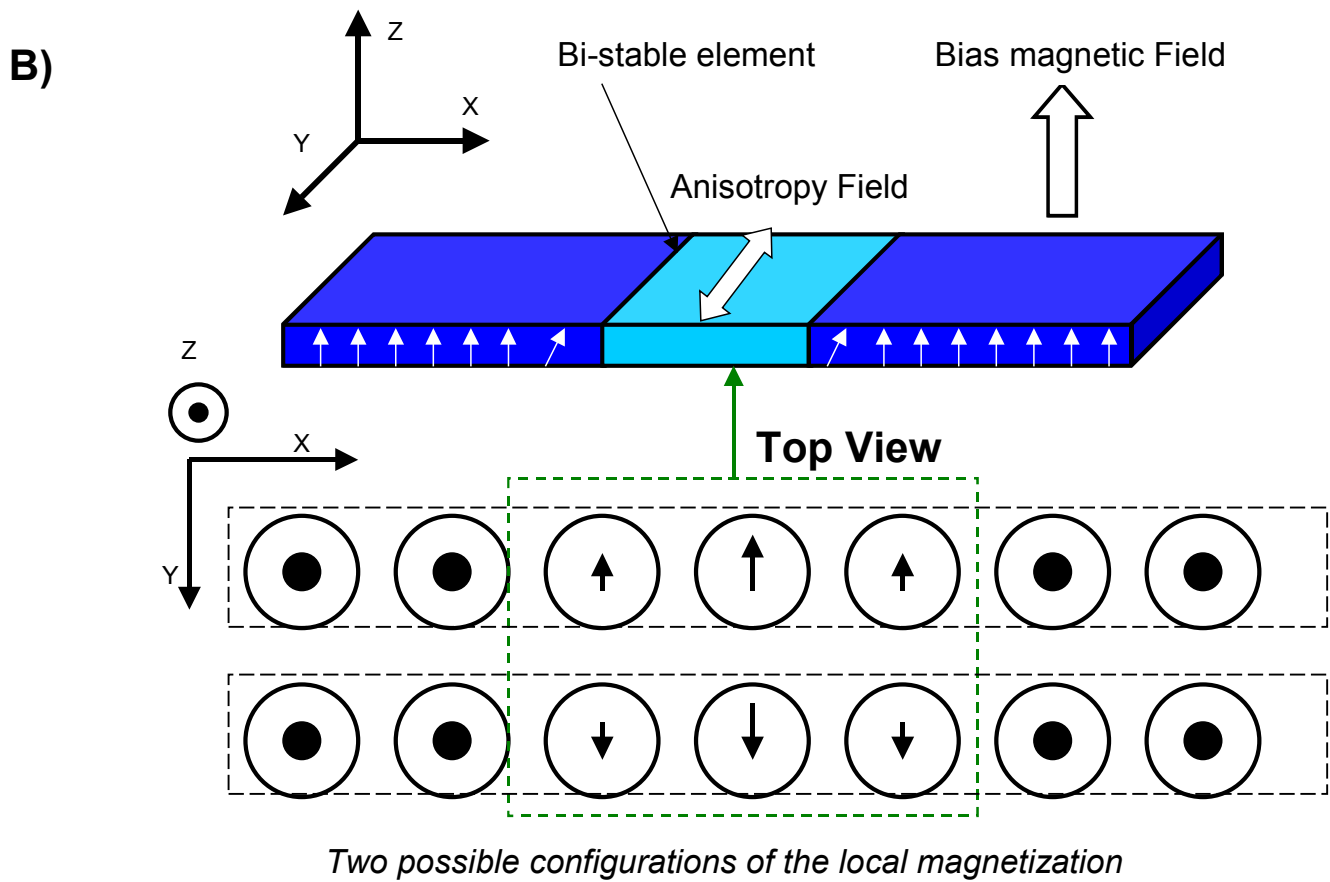

*Two possible configurations of the local magnetization*

**Fig.2**

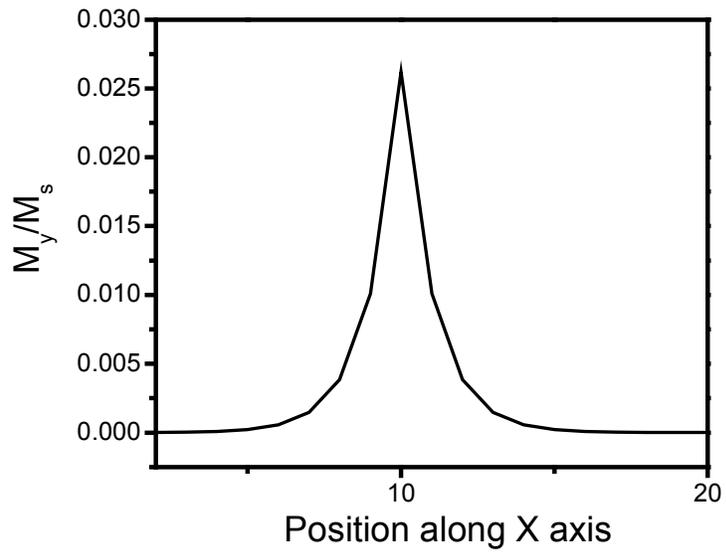
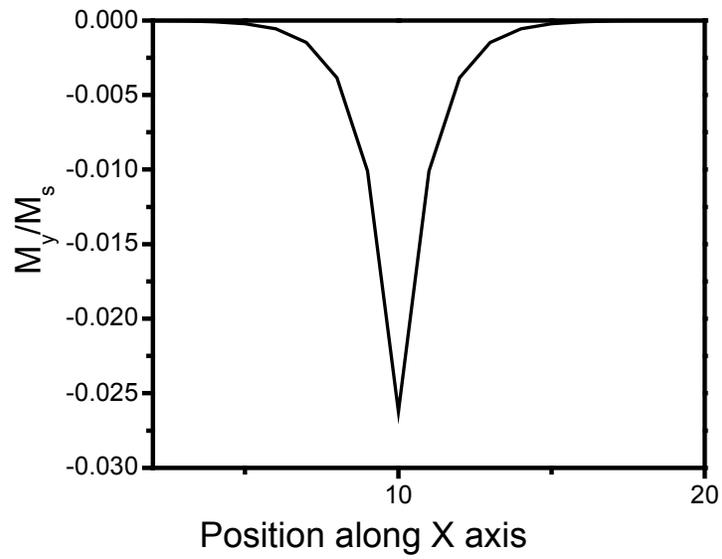

**Fig.3**

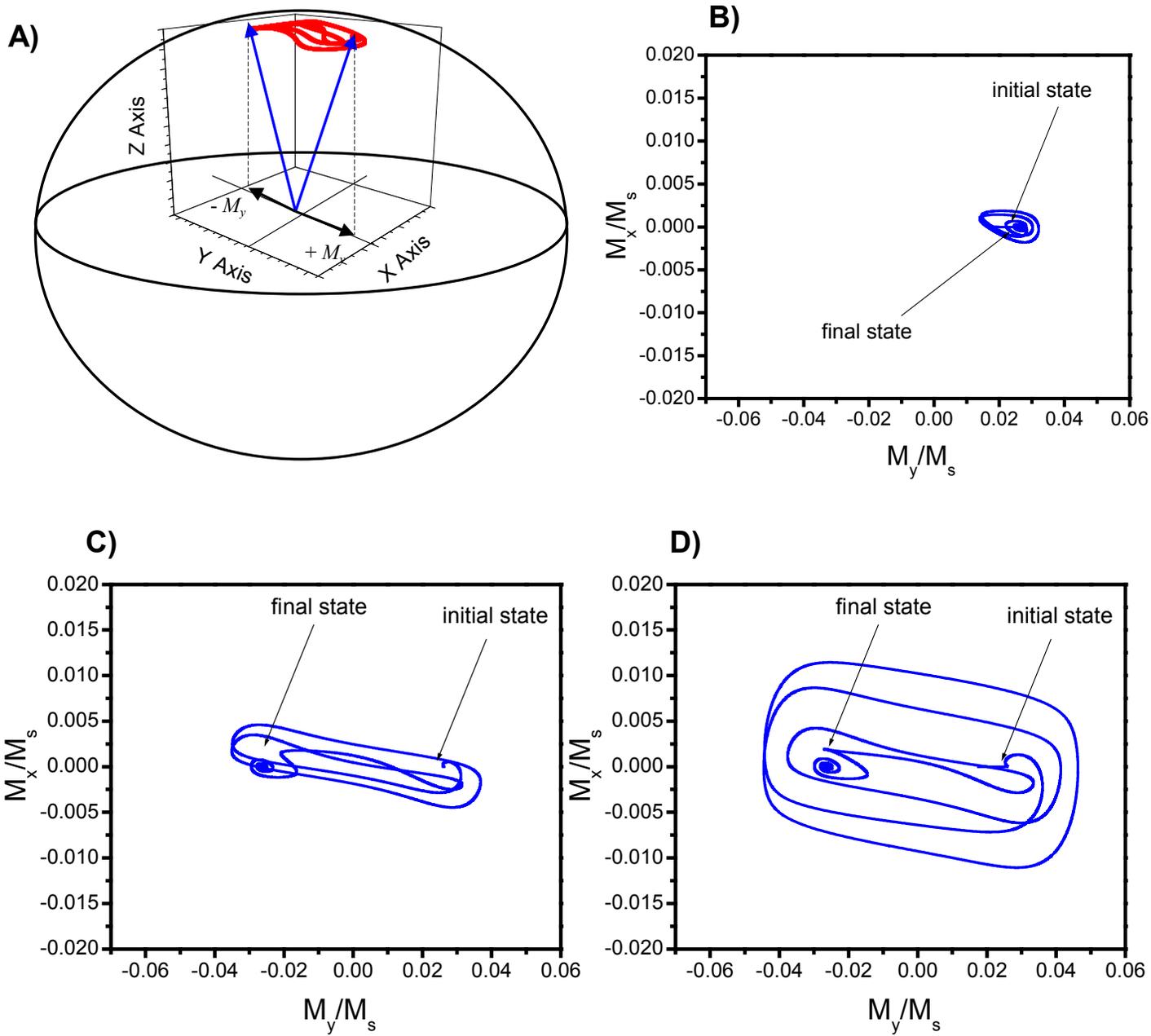

**Fig.4**

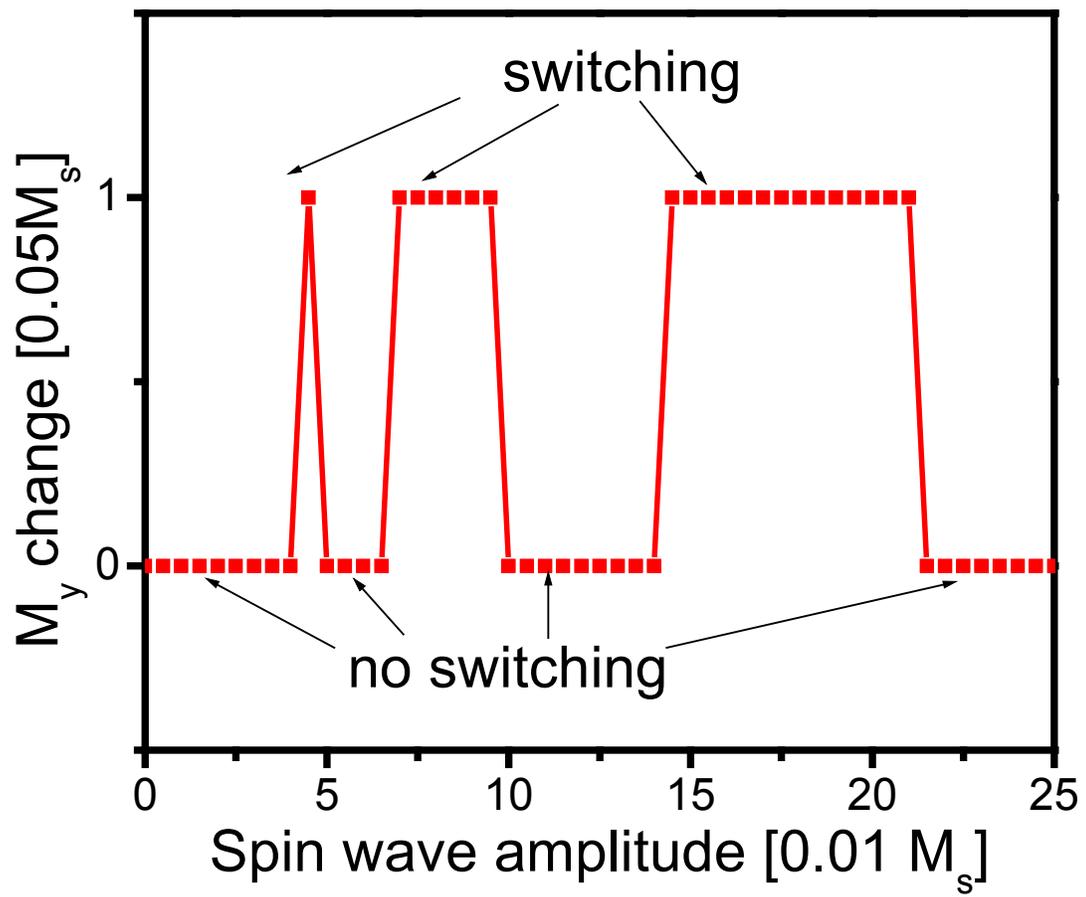

**Fig.5**

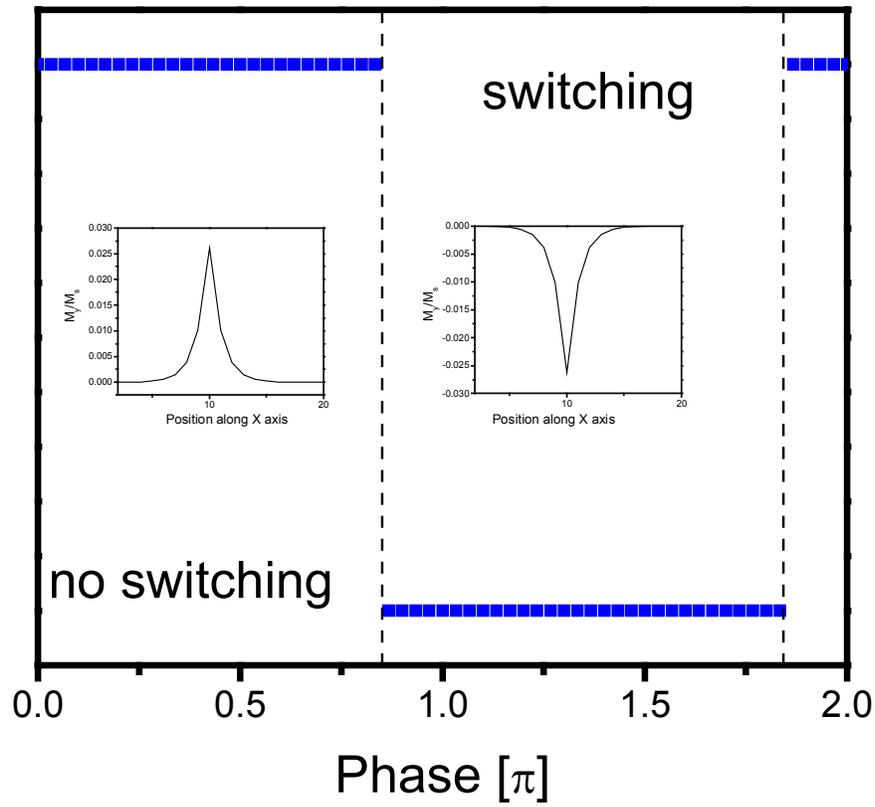

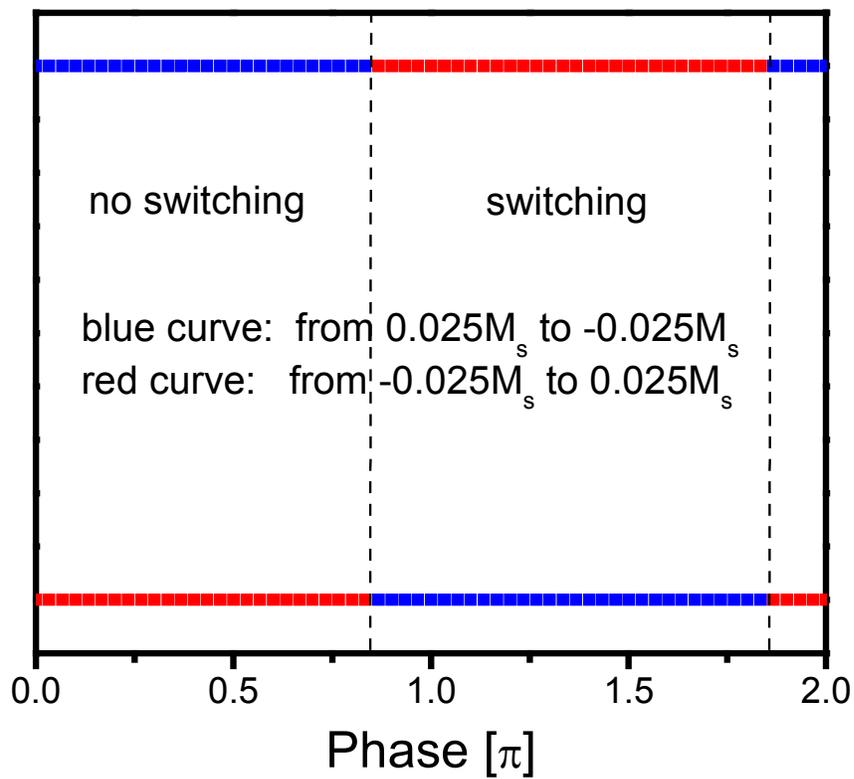

**Fig.6**

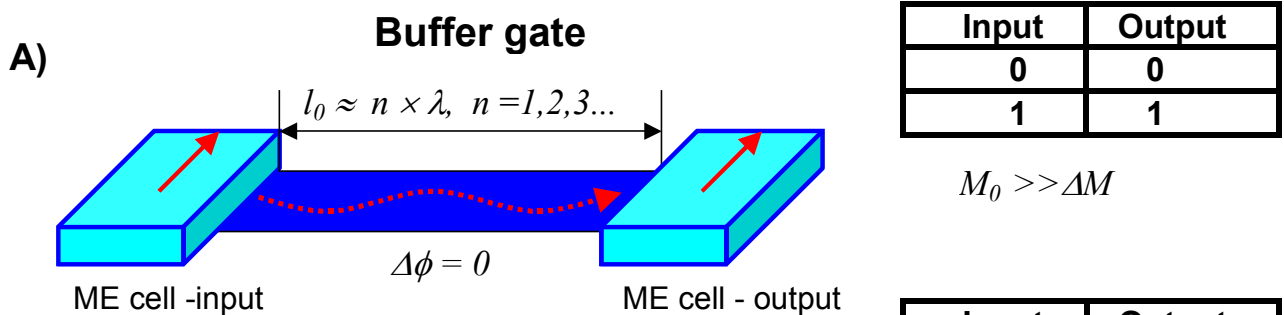
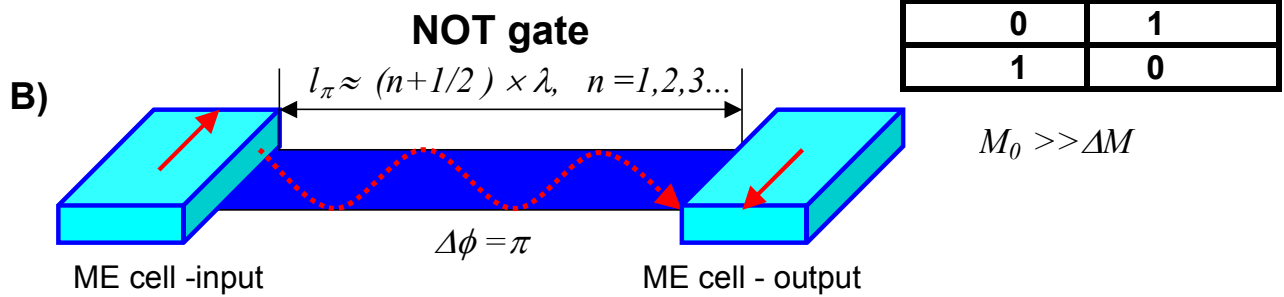
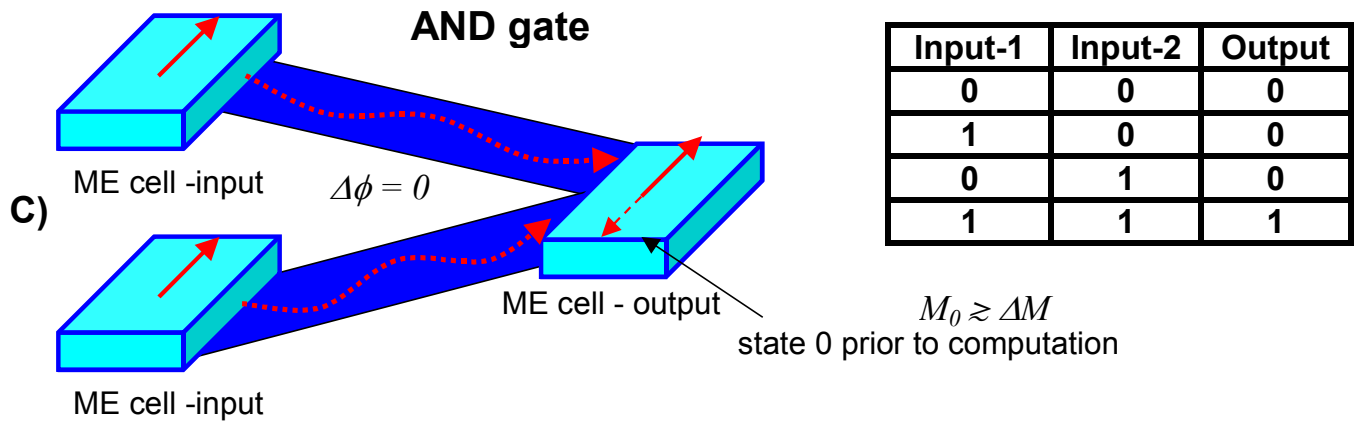
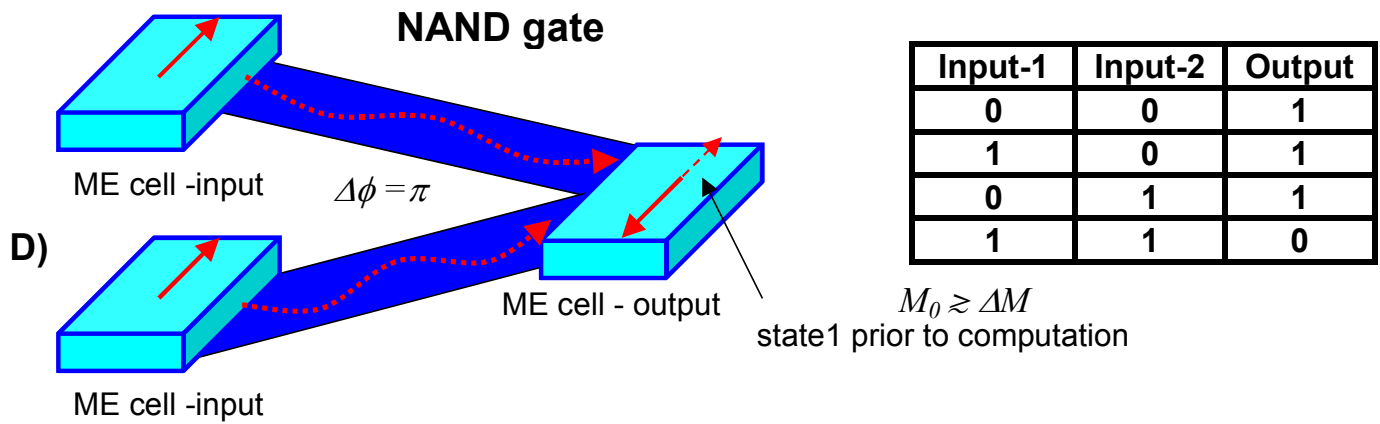

**Fig.7**

# MAJ gate

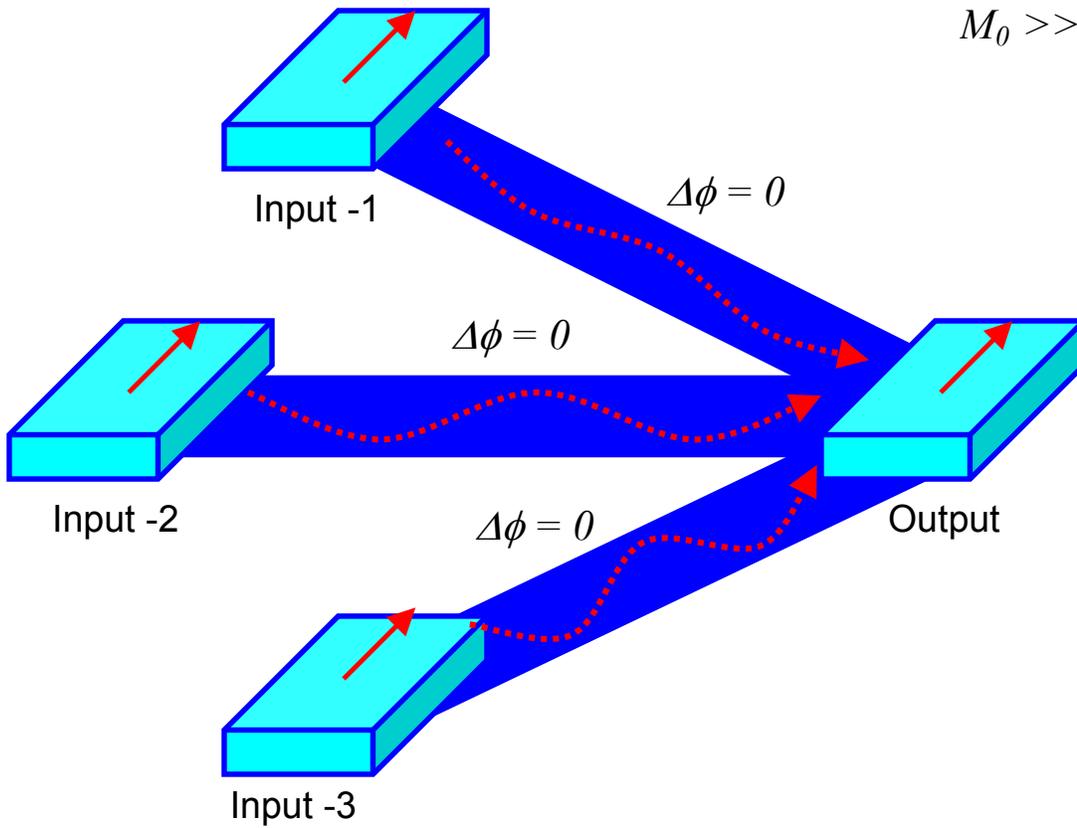

$M_0 >> \Delta M$

$\Delta\phi = 0$

| Input -1 | Input -2 | Input -3 | Output |
|---|---|---|---|
| 0 | 0 | 0 | 0 |
| 0 | 0 | 1 | 0 |
| 0 | 1 | 0 | 0 |
| 0 | 1 | 1 | 1 |
| 1 | 0 | 0 | 0 |
| 1 | 0 | 1 | 1 |
| 1 | 1 | 0 | 1 |
| 1 | 1 | 1 | 1 |

**Fig. 8**

# XOR gate

| Input-1 | Input-2 | Output |
|---------|---------|--------|
| 0 | 0 | 0 |
| 1 | 0 | 1 |
| 0 | 1 | 1 |
| 1 | 1 | 0 |

$\Delta\phi = \pi$

pinned layer

Input -1

Input -2

Output

$\Delta\phi = 0$

**Fig. 9**

# MOD gate

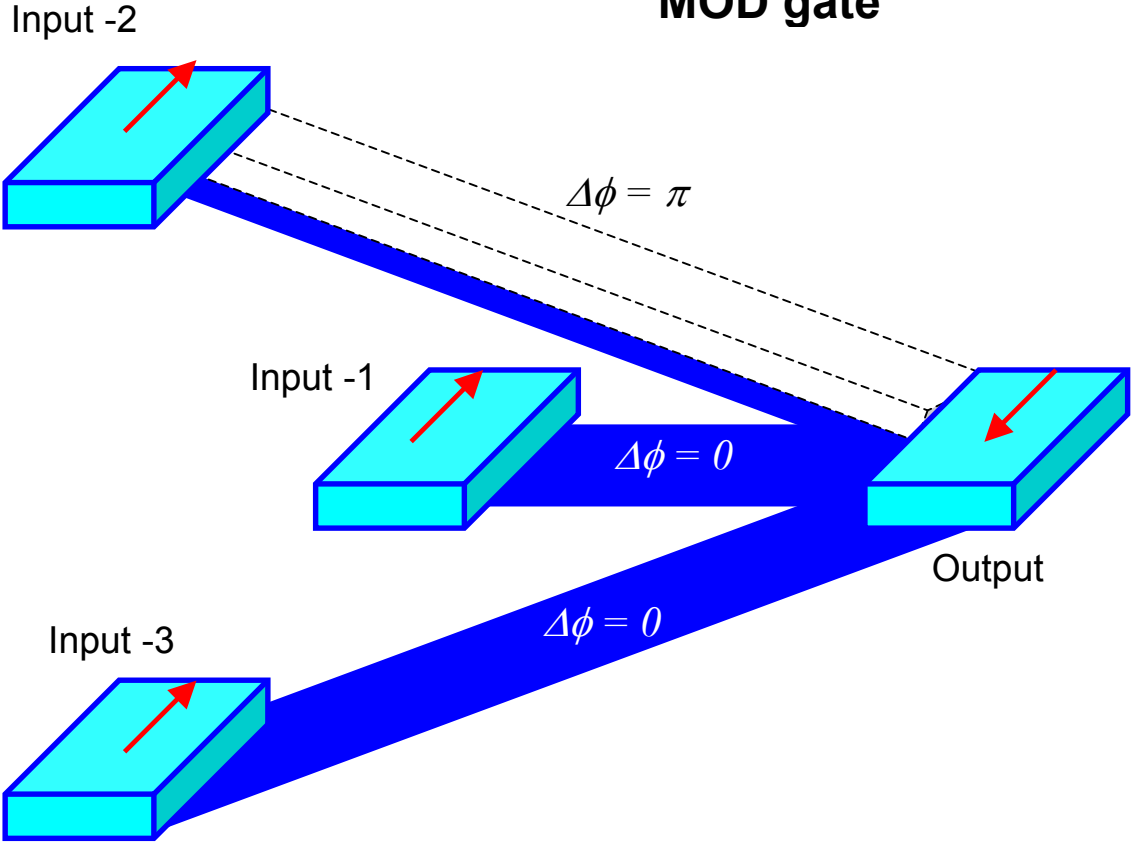

| Input - 1 | Input - 2 | Input - 3 | Output' | Output |
|---|---|---|---|---|
| 0 | 0 | 0 | 0 | 0 |
| 0 | 0 | 1 | 0 | 1 |
| 0 | 1 | 0 | 0 | 1 |
| 0 | 1 | 1 | 0 | 0 |
| 1 | 0 | 0 | 1 | 1 |
| 1 | 0 | 1 | 1 | 0 |
| 1 | 1 | 0 | 1 | 0 |
| 1 | 1 | 1 | 1 | 1 |

**Fig. 10**

## Full Adder Circuit

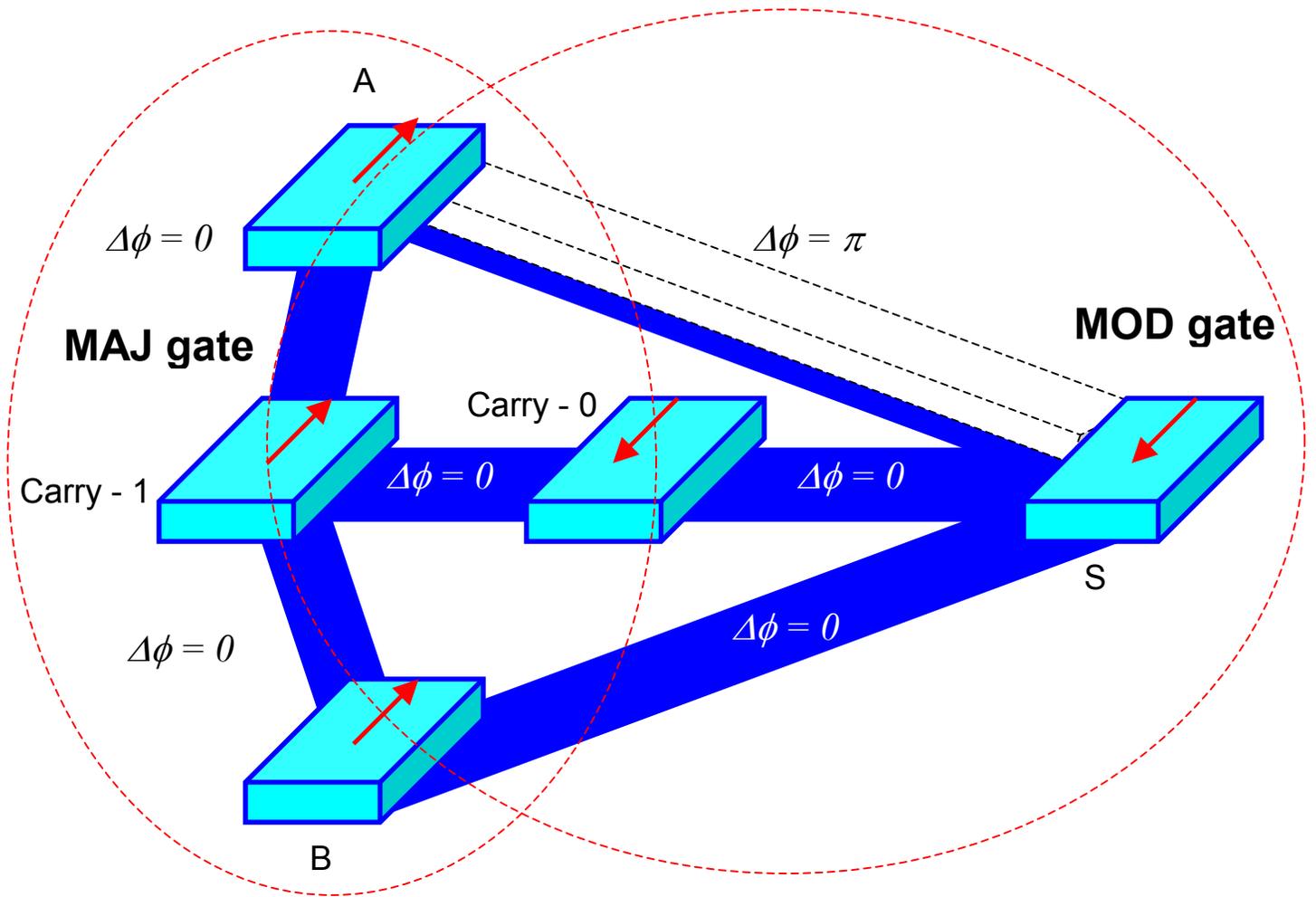

| Input - A | Input - B | Carry - 0 | Carry - 1 | Output - S |
|---|---|---|---|---|
| 0 | 0 | 0 | 0 | 0 |
| 0 | 0 | 1 | 0 | 1 |
| 0 | 1 | 0 | 0 | 1 |
| 0 | 1 | 1 | 1 | 0 |
| 1 | 0 | 0 | 0 | 1 |
| 1 | 0 | 1 | 1 | 0 |
| 1 | 1 | 0 | 1 | 0 |
| 1 | 1 | 1 | 1 | 1 |

**Fig. 11**

|  | **45nm CMOS** | **32nm CMOS** | **λ=45nm** | **λ=32nm** |
|---|---|---|---|---|
| **Area** | 6.4 μm$^2$ | 3.2 μm$^2$ | 0.05 μm$^2$ | 0.026 μm$^2$ |
| **Time Delay** | 12 ps | 10 ps | 13.5 ps | 9.6 ps |
| **Functional Throughput** | 1.3×10$^9$ Ops/[ns cm$^2$] | 3.1×10$^9$ Ops/[ns cm$^2$] | 1.48×10$^{11}$ Ops/[ns cm$^2$] | 4.0×10$^{11}$ Ops/[ns cm$^2$] |
| **Energy per operation** | 12fJ | 10fJ | 24aJ | 24aJ |
| **Static Power** | > 70nW | > 70nW | 0nW | 0nW |

**Fig. 12**